*Model of skyscraper evacuation with the use of space symmetry and fluid dynamic approximation,*

W. Sikora, J. Malinowski, A. Kupczak

Faculty of Physics and Applied Computer Science, AGH - University of Science
and Technology, Al. Mickiewicza 30, 30-059 Krakow, Poland

======================================================

The simulation of evacuation of pedestrians from skyscraper is a situation where the symmetry analysis method and equations of fluid dynamics finds to be very useful. When applied, they strongly reduce the number of free parameters used in simulations and in such a way speed up the calculations and make them easier to manage by the programmer and what is even more important, they can give a fresh insight into a problem of evacuation and help with incorporation of „Ambient Intelligent Devices" into future real buildings. We have analyzed various, simplified, cases of evacuation from skyscraper by employing improved „Social Force Model". For each of them we obtained the average force acting on the pedestrian as a function of the evacuation time. The results clearly show that both methods mentioned above, can be successfully implemented in the simulation process and return with satisfactory conclusions.

======================================================

Analytical models of socio technical systems which use sets of local parameters may be significantly simplified, when the number of parameters is reduced to smallest number of relevant ones. The symmetry analysis method (SAM) offers such a possibility (described in [1]) because the behavior of social systems under the action of some external conditions may be regarded as similar to the behavior of solid states under the action of temperature or external electric or magnetic fields. Both systems are complex, containing many interacting elements, and are realized in strictly defined spaces. Very often these spaces are strongly restricted, and because of these restrictions not all types of evolutions of these systems are allowed. When these spaces are symmetrical (crystals are good examples of such situation) the symmetry considerations conducting in the frame of theory of groups and their representations are able to predict the types of behavior of the systems, which are permitted by the symmetry of these spaces. The SAM [1] was successfully used for many years to significantly simplify the descriptions of different type of phase transitions in crystals. The discussion of coexistence of different types of system behavior, leading to different properties of the system are based on the assumption, that different functions describing these properties of the system should have the same symmetry. From the theory of representation we know, that these functions should belong to the same irreducible representation of space symmetry group. It gives the powerful instrument to discussions of complex behavior of the systems, because we have the possibility to extract all limitations of free parameters needed for description of the models. The local parameters such as the resultant force acting on each pedestrian and their velocities are related by the symmetry conditions and expressed by the smallest number of relevant coefficients.

The SAM was applied to investigate what is the influence of space symmetry on the quality of crowd evacuation plans and number of model parameters used in crowd behavior simulation for a football stadium and one floor of multi-floor skyscraper [2],[3]. The simulations were performed within the Social Force Model designed by Helbing in 2000 [4], developed by J. Malinowski. He introduced in the program the symmetry adapted vector field calculated by the symmetry analysis method as one of the forces acting on the pedestrian during evacuation and the possibility of evacuation from complicated space [3]. The last, new version of the program offers possibility of introducing individual, physical parameters of the agents. (Appendix). The results for multi-floor skyscraper evacuation were obtained by using the symmetry of one floor [2] and the translational symmetry between different floors. Taken into account the translational symmetry of skyscraper the discussion of evacuation is limited only to one "cell" of the structure – it means only to two floors connected by staircases ( different geometry depending on the building architecture is possible).

Under the assumption that the velocity of agents at the staircase $v_s$ is constant, the constant density of agents which ensures comfortable evacuation may be got by introducing time shift between the start of evacuation from different floors. The optimal value of this time shift may be calculated from continuity equation as the function of building geometry parameters and social system parameters (for example the number of pedestrian on each floor, their possible velocities…) .

The time shift between the beginning of evacuation of first floor and the next floor, which guarantees the smooth evacuation with constant density of evacuated individuals, should be:

$$\Delta t = t_f - t_s \quad (1)$$

where $t_f$ is the time of evacuation of one floor, $t_s$ is the time needed for leaving the staircase which connect two floors. The evaluation of $t_f$ for a given number of pedestrians located on a specific floor with symmetry considerations, was discussed in [3].

The continuity equation and geometry of the building leads to the relations:

$$v_f * c_f = v_s * c_s ; \quad v_s = v_f * c_f / c_s \quad (2)$$

where $v_f$ and $v_s$ are speeds of agents at the exit from the floor and at the staircase correspondingly, $c_f$ is the width of exit and $c_s$ is the width of the staircase.

The time $t_s$ needed for covering the staircase length $l_s$ is given as:

$$v_s * t_s = l_s ; \quad t_s = l_s / v_s; \quad t_s = l_s * c_s / v_f * c_f \quad (3)$$

then

$$\Delta t = t_f - t_s = t_f - l_s * c_s / v_f * c_f \quad (4)$$

Evaluation of this time shift and sending the information about the accident to given floor at proper time requires high quality ambient intelligence device (AID), which is able to follow and keep in memory the number of pedestrian on given skyscraper floor, and to use this information for determination of floor evacuation, and appropriate time shift.

In this work the simulations of evacuation with calculated time shift are realized in skyscrapers with different staircases. In each case the average force acting on the agent as the function of evacuation time is calculated and compared with similar results got from simulations when the evacuation starts from different floors at the same time. The simplest models, with the same width of exits and staircases are investigated here. The results are presented at the figures showed below.

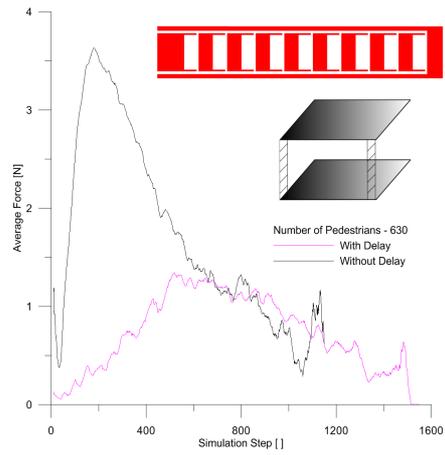

Fig.1 Average force per pedestrian as a function of evacuation time – fire escape ladder

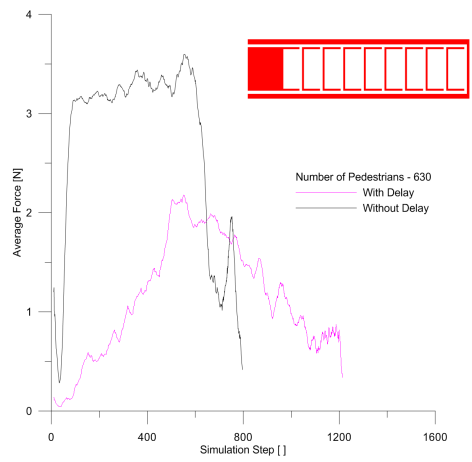

Fig.2 Average force per pedestrian as a function of evacuation time – fire escape ladder

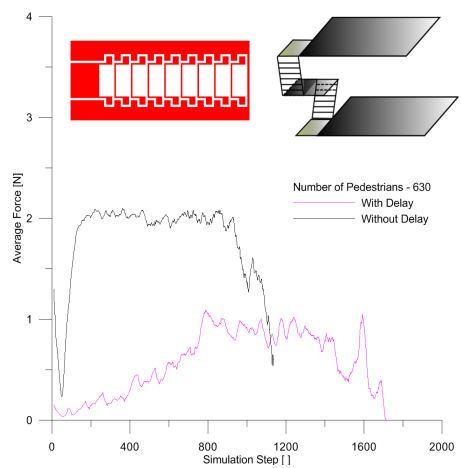

Fig.3 Average force per pedestrian as a function of evacuation time – standard staircase

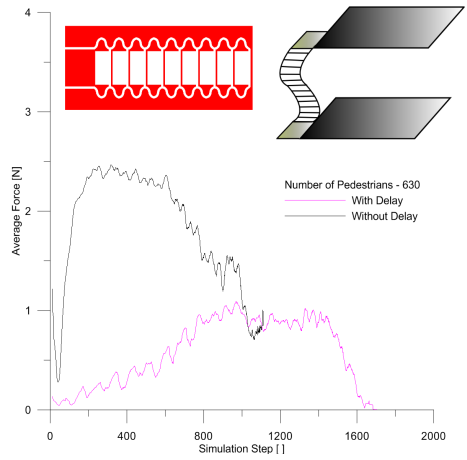

Fig.4 Average force per pedestrian as a function of evacuation time- helical staircase

Fig.1 and Fig.2 show two types of evacuation routes (the difference lies in the length of escape path) which can often be found at the back of the building – fire escape ladder. On the other hand in the Fig.3 we deal with a standard staircase that can be seen in the block of flats as well as in dormitory. The helical staircase in Fig.4 differs from the one in Fig. 3 only in the shape of the stairs.

As it can be seen in the presented figures imposing the time shift between start of the evacuation from different floors makes the time of the evacuation longer, but significantly reduces the average force acting on the pedestrian. Therefore there is a visible need to put more effort in such investigations, so that the intelligent ambient devices found their places in future building industry.

Simulations with different evacuation parameters will be realized in the future.

Appendix

The evacuation of the skyscraper is simulated by using the Helbing "social forces" model [5], which was taken as a starting point to model presented in this paper.

$$m_i \frac{d\vec{v}_i}{dt} = m_i \frac{v_i^0 \vec{e}_i^0(t) - \vec{v}_i(t)}{\tau} + \sum_{j(\neq i)} \vec{f}_{ij} + \sum_W \vec{f}_{iW} \quad \text{(A1)}$$

Equation A1 on right side has a sum of three forces acting on human during evacuation simulation. Second and third one describe sum of repel forces from all other mans and sum of repel forces from all walls and all obstacles inside the evacuated space. First force is the most interesting at this stage. In general it describes in which direction each person should go. In simple case like empty room with some doors, this force can be easily calculated as a sum of attract forces from all doors. Here is also the place for the implementation of vector field calculated by symmetry considerations. This assumption is inadequate with more complicated geometry of the building. Figure (A1) shows the example when such approach leads to situation when pedestrian will stick for eternity in dead end.

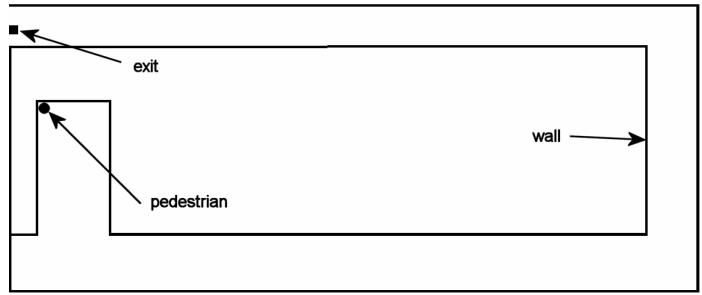

FigA1. The „dead-end".

In approach presented in this paper floor space was divided into cells (this got nothing in common with "cellular automata" approach to such kind of simulations!). Every cell can contain obstacle, free space or exit. Every cell that contain free space have vector of desirable velocity connected with it, calculated according to schema described below. Any person that stands inside particular cell takes this vector as his desirable velocity. Unlike in cellular automata approach in this approach "cell" can contain one or more persons and any direction of velocity. Single cell size depends on building geometry only. To make a simulation of an evacuation of single room, small number (like 7x6 for example) of cells is more than enough. Figure (A2) show example of such table presenting the final field of velocities. These vectors can be calculated using "ray casting" method as shown in paper [6].

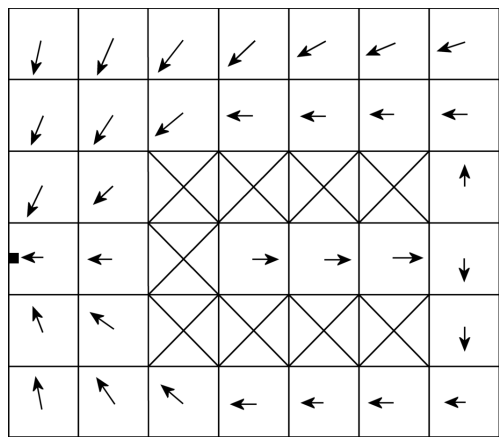

FigA2. Final field of velocities

Acknowledgement

This work is partially supported by EU program SOCIONICAL (FP7 No 231288).